\documentclass[aps, twocolumn, prl, longbibliography]{revtex4-1}
\usepackage{amsfonts}
\usepackage{amsmath}
\usepackage{amssymb}
\usepackage{mathptmx} 
\usepackage{graphicx}
\usepackage{bm}
\usepackage{color}
\usepackage[colorlinks=true,linkcolor=blue,anchorcolor=red,citecolor=blue,urlcolor=blue]{hyperref}
\usepackage{ulem}
\usepackage{natbib}

\begin{document}
\title{Screening effects of superlattice doping on the mobility of GaAs two-dimensional electron system revealed by in-situ gate control}
\author{T. Akiho and K. Muraki}
\affiliation{NTT Basic Research Laboratories, NTT Corporation, 3-1 Morinosato-Wakamiya, Atsugi 243-0198, Japan}
\keywords{one two three}
\pacs{PACS number}
\date{\today}

\begin{abstract}
We investigate the screening effects of excess electrons in the doped layer on the mobility of a GaAs two-dimensional electron system (2DES) with a modern architecture using short-period superlattice (SL) doping. By controlling the density of excess electrons in the SL with a top gate while keeping the 2DES density constant with a back gate, we are able to compare 2DESs with the same density but different degrees of screening using one sample. Using a field-penetration technique and circuit-model analysis, we determine the density of states and excess electron density in the SL, quantities directly linked to the screening capability. The obtained relation between mobility and excess electron density is consistent with the theory taking into account the screening by the excess electrons in the SL. The quantum lifetime determined from Shubnikov-de Haas oscillations is much lower than expected from theory and did not show a discernible change with excess electron density.
\end{abstract}
\maketitle
\section{I. INTRODUCTION}
High-mobility two-dimensional electron systems (2DESs) in AlGaAs/GaAs heterostructures are the basic platform to test new concepts and study emergent phenomena in low-dimensional systems.
The modulation doping technique that separates the channel and the doping layer for carrier supply \cite{Pfeiffer2003,Umansky2009,Gardner2016,Manfra2014,Chung2020} and advances in molecular-beam epitaxy that enable the residual impurity concentration to be decreased are the key ingredients in realizing clean 2DESs.
Over the years, improvements in sample quality, manifested as higher mobility, have led to the discovery of new transport phenomena \cite{RLWillett1988,Goldman1990,Lilly1999,Du1999,Suen1992} and correlated phases including the fractional quantum Hall effects (FQHEs) \cite{Tsui1982,Willett1987}.
However, it has recently been recognized that not only mobility but also the screening of the long-range disorder potential caused by modulation doping is essential for the observation of fragile FQHEs such as the one at an even-denominator Landau-level filling factor $\nu = 5/2$ \cite{Umansky2009,Pan2011,Gamez2013}.
Specifically, modulation doping in an AlAs/GaAs/AlAs short-period superlattice (SL) \cite{Friedland1996} or low-$x$ Al$_x$Ga$_{1-x}$As ($x = 0.24$--$0.25$) alloy \cite{Gamez2013} has been shown to be effective, where the electrons in the doped layer delocalize and screen the Coulomb potential from ionized donors.

\begin{figure}[pb]
\includegraphics[scale=1.08]{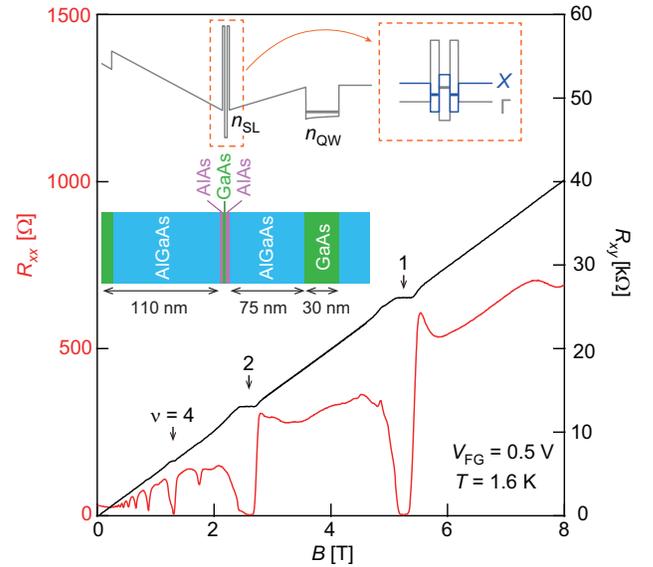}\caption{
Magnetotransport properties, $R_{xx}$ (red) and $R_{xy}$ (black), of the Hall bar measured at 1.6 K. Insets show the schematic layer structure of the sample and $\Gamma$-conduction band edge profile. The X- and $\Gamma$- conduction band edge profiles near the AlAs/GaAs/AlAs superlattice are shown on an enlarged scale in the dashed box.
}
\label{Fig1}
\end{figure}

The concept of SL doping where a $\delta$-doped donor layer is located within a narrow GaAs layer flanked by narrow AlAs layers was originally introduced by Friedland \textit{et al.} to reduce remote-impurity scattering and thereby enhance mobility \cite{Friedland1996}.
In the AlAs/GaAs/AlAs SL, the energy level of the X-band formed in the AlAs layers is lower than that of the $\Gamma$-band formed in the GaAs layer (inset of Fig.~\ref{Fig1}).
Consequently, mobile electrons supplied from the donor layer accumulate not only in the GaAs quantum well (QW) several tens of nanometers away but also in the neighboring AlAs layers.
The SL doping technique was later applied to ultra-high-quality samples with mobility exceeding $10 \times 10^6$~cm$^{2}$/Vs, where its impact on the FQHEs has been demonstrated \cite{Pfeiffer2003,Umansky2009,Gardner2016,Manfra2014,Chung2020}.
Recently, effects of excess electrons in the SL on mobility and the quantum scattering lifetime have been studied theoretically \cite{SammonChe2018,SammonZu2018}. Experimentally, gating of samples with SL doping has been attempted to examine the influences of the parallel conducting layer on mobility \cite{Friedland1996,Rossler2010,Dmitriev2012,Peters2017}.
However, uncontrollable charge redistribution and hysteresis that accompany the SL doping \cite{Rossler2010} have made it difficult to extract quantitative information such as the density of excess electrons in the SL.

In this study, we vary the excess electron density in the SL in a controlled manner by appropriately choosing the temperature at which the gate voltage is swept.
This enabled the in situ control of disorder screening.
We determined the electron density in the SL as a function of gate voltage by using a field-penetration technique \cite{Eisenstein1994} and circuit-model analysis, which also allowed us to estimate the quantum capacitance, or the density of states (DOS), in the SL.
We show that the obtained relation between mobility and excess electron density can only be explained by theory taking into account the screening by excess electrons.

\section{II. EXPERIMENT and ANALYSIS}
\subsection{A. Sample characterization}
The sample consisted of a 30-nm-wide GaAs QW sandwiched between Al$_{0.27}$Ga$_{0.73}$As barriers, grown on an $n$-type GaAs (001) substrate. The QW, with its center located 207 nm below the surface, was modulation doped on one side, with Si $\delta$-doping ([Si] = $1 \times 10^{16}$~m$^{-2}$) at the center of the AlAs/GaAs/AlAs (2 nm/3 nm/2 nm) SL located 75 nm above the QW (inset of Fig.~\ref{Fig1}) \cite{Gardner2016}. The Si $\delta$-doping in the thin GaAs layer provides mobile electrons not only in the QW 75-nm away but also in the neighboring AlAs layers \cite{Umansky2009,Gardner2016,Manfra2014,Chung2020,Friedland1996}.
The mobile electrons in the AlAs layers provide screening of the disorder potential created by the ionized Si donors. The wafer was processed into a 120-$\mu$m-wide Hall bar with voltage-probe distance of 100 $\mu$m and fitted with a Ti/Au front gate. The $n$-type substrate was used as a back gate. Measurements were done at temperatures of $0.27$--$4.3$ K using a standard lock-in technique.

Figure~\ref{Fig1} shows the magnetotransport of the sample measured at 1.6 K. Here, we show data taken at a positive front gate voltage ($V_\text{FG}$) of 0.5 V, which is supposed to increase the density of mobile electrons in the SL. Interestingly, despite the presence of mobile electrons in the SL, there are integer quantum Hall effects at Landau-level filling factor $\nu = 1, 2$, and $4$, where the longitudinal resistance ($R_{xx}$) drops to zero and the Hall resistance ($R_{xy}$) is quantized. From the Shubnikov-de Haas oscillations, we obtained a sheet carrier density of $1.25 \times 10^{15}$~m$^{-2}$, which agrees within 3\% with the value deduced from the slope of $R_{xy}$. This suggests that, even if the SL contains conduction electrons, they apparently do not contribute to transport. We confirmed similar results for $V_\text{FG}$ up to 0.8 V.

\begin{figure}[ptb]
\includegraphics[scale=1.08]{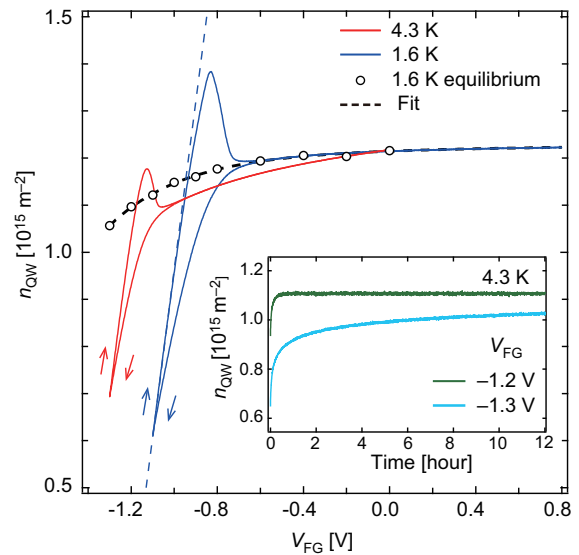}\caption{
$V_\text{FG}$ dependence of $n_\text{QW}$ obtained from $R_{xy}$ at $B = \pm0.2$ T. Blue and red curves show $n_\text{QW}$ obtained by sweeping $V_\text{FG}$ at 1.6 and 4.3 K, respectively. The dashed blue line is a linear fit for the up sweep. Open circles show $n_\text{QW}$ in the equilibrium state (see main text for details). Black dashed line is a fit using a double exponential function. Inset shows the time evolution of $n_\text{QW}$ at $V_\text{FG} = -1.2$ and $-1.3$ V at 4.3 K.}
\label{Fig2}
\end{figure}
Figure~\ref{Fig2} shows the $V_\text{FG}$ dependence of the sheet carrier density deduced from $R_{xy}$ at 0.2 T.
The blue solid curve was obtained by sweeping $V_\text{FG}$ at 1.6 K.
As shown above, at 1.6 K the measured $R_{xy}$ reflects exclusively the carrier density in the quantum well ($n_\text{QW}$), and not that in the SL.
As we decreased $V_\text{FG}$ from 0.8 V, $n_\text{QW}$ remained almost constant for $-0.7$ $< V_\text{FG} <$  $0.8$ V (region I).
This is consistent with the expectation that the SL contains mobile electrons, which screen the electric field from the front gate and thereby suppress its effect on $n_\text{QW}$.
A distinct change in $n_\text{QW}$ occurs only at $V_\text{FG} < -0.7$ V (region II), indicating that the screening capability of the SL is significantly decreased in region II.
Upon increasing $V_\text{FG}$, we observed a pronounced hysteresis in region II, where $n_\text{QW}$ changed at a faster rate, with an overshoot near the boundary with region I.
The rate of change d$n_\text{QW}$/d$V_\text{FG} = 3.5 \times 10^{15}$ m$^{-2}$V$^{-1}$ for the up sweep, shown by the blue dashed line, was consistent with the geometrical capacitance between the front gate and the center of the QW calculated from the distance (207 nm) and the permittivity of AlGaAs ($\epsilon = 13$).
This implies that, upon increasing $V_\text{FG}$ in region II, electrons accumulate only in the QW, resulting in a metastable state in which the QW (SL) is overpopulated (underpopulated) with respect to its equilibrium density.
We conjecture that this results from the difficulty to inject charge into the SL once it becomes close to depletion and poorly conducting.
The red curve in Fig.~\ref{Fig2}, obtained by sweeping $V_\text{FG}$ at 4.3 K, shows similar behavior, while the boundary between regions I and II shifts to a more negative $V_\text{FG}$, with a smaller overshoot in the up sweep.

Turning our attention to region I, we notice that $n_\text{QW}$ is not constant, but varies slightly with $V_\text{FG}$.
This indicates that part of the electric field from the gate penetrates the SL populated with electrons.
This is reasonable, as the SL is not a perfect metal; it has only a finite DOS, that is, a finite screening capability.
In turn, by analyzing the change in $n_\text{QW}$ with $V_\text{FG}$ as shown later, we can quantify the DOS, and hence the screening capability, of the SL. (See Ref. \cite{Eisenstein1994} for the principle of this field-penetration technique.)
\begin{figure}[ptb]
\includegraphics[scale=1.05]{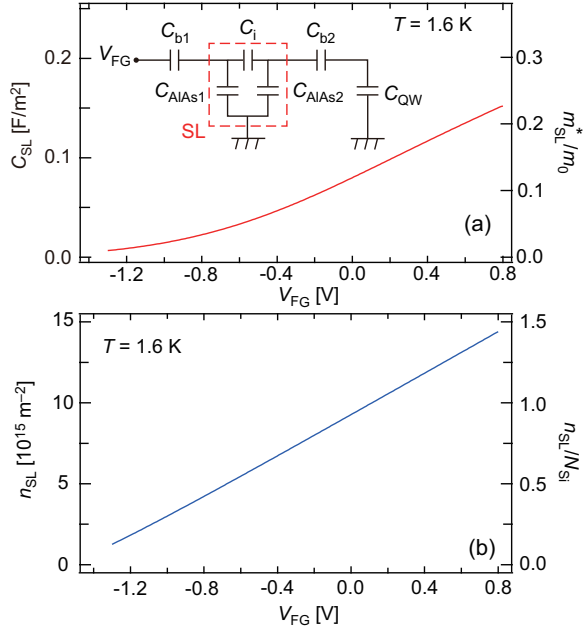}\caption{
red(a) Quantum capacitance $C_\text{SL}$ of the AlAs layers comprising the SL calculated from the fitting of $n_\text{QW}$ in the equilibrium state. The right axis shows the corresponding effective mass for electrons in the SL. Inset shows the equivalent circuit model used to calculate $C_\text{SL}$. We assume $C_\text{AlAs1} = C_\text{AlAs2} \equiv C_\text{SL}$ (see main text for details). (b) Calculated $V_\text{FG}$ dependence of $n_\text{SL}$ red($= n_\text{AlAs1}+n_\text{AlAs2}$).
The right axis indicates $n_\text{SL}$ normalized by the doping density $N_\text{Si} = 1 \times 10^{16}$~m$^{-2}$.
}
\label{Fig3}
\end{figure}

Even though we used a very slow sweep rate of 0.67 mV/sec to set $V_\text{FG}$, in region II $n_\text{QW}$ gradually increased on a scale of several minutes to several tens of hours after $V_\text{FG}$ was set at a constant value (inset of Fig.~\ref{Fig2}).
This was the case even for down sweeps for which the system is closer to equilibrium.
Similar temporal behavior was reported in Ref. \cite{Rossler2010}. The transient time increased with decreasing temperature and decreasing $V_\text{FG}$.
At $V_\text{FG} = -1.3$ V, equilibrium was not reached even after a few days at 4.3 K. We used the following method to determine the equilibrium value of $n_\text{QW}$ at each $V_\text{FG}$, which is essential for evaluating the screening effect.
First, we set $V_\text{FG}$ at 4.3 K to facilitate the equilibration and waited until $n_\text{QW}$ reached a steady value.
Then, we decreased the temperature to 1.6 K and determined $n_\text{QW}$ from the low-field $R_{xy}$. By repeating this process for different $V_\text{FG}$, we obtained $n_\text{QW}$ as a function of $V_\text{FG}$, which we plot as open circles in Fig.~\ref{Fig2}.
At $V_\text{FG} = 0$ V, we obtained the same $n_\text{QW}$ value as that for the $V_\text{FG}$ sweep.
However, the difference between the two methods became significant at lower $V_\text{FG}$.
We therefore employed the data obtained by the equilibration method for $V_\text{FG} \leq$ 0 V and those of the $V_\text{FG}$ sweep for $V_\text{FG} > 0$ V and fit them using a double exponential function (the black dashed line in Fig.~\ref{Fig2}).

\subsection{B. Circuit model}
To deduce the excess electron density in the SL and thereby quantitatively characterize the screening effect, we analyzed the charge equilibration among the front gate, redtwo AlAs layers [AlAs1(2)] comprising the SL, and QW using the circuit model shown in the inset of Fig.~\ref{Fig3}(a).
In addition to the geometrical capacitances between the neighboring elements among these ($C_\text{b1}$, $C_\text{i}$, and $C_\text{b2}$), the model contains the quantum capacitances of the QW ($C_\text{QW}$) and the AlAs layers [$C_\text{AlAs1(2)}$].
The quantum capacitance is expressed as $C_\alpha = e^2D_\alpha$, where $D_\alpha = g_\alpha m^\star_\alpha$/$2\pi\hbar^2$ is the DOS ($m^\star_\alpha$ is the electron effective mass, $\alpha$ denotes QW or AlAs1(2), $e$ is the elementary charge, $g_\alpha$ is the degeneracy, and $\hbar = h$/2$\pi$ is the reduced Planck constant).
$C_\text{b1}$, $C_\text{i}$, and $C_\text{b2}$ are calculated from the layer thicknesses and permittivity and $C_\text{QW}$ is known from the effective mass $m^\star_\text{QW} = 0.067m_0$ of GaAs ($m_0$ is the electron mass in vacuum) and the twofold spin degeneracy.
This leaves $C_\text{AlAs1(2)}$ the only unknown parameters in the model.
For the model to be solvable, we need to assume that the two AlAs layers have the same density of states at the Fermi level, that is, $C_\text{AlAs1} = C_\text{AlAs2}$  ($\equiv C_\text{SL}$).
We confirmed this assumption to be acceptable by noting that the calculated chemical potential difference between the two AlAs layers (up to 7 meV) was smaller than the disorder-broadened tail (a few tens of meV).

We calculate $\text{d}n_\text{QW}/\text{d}V_\text{FG}$ as a function of $V_\text{FG}$ using the equilibrium relation between $n_\text{QW}$ and $V_\text{FG}$ obtained above.
By numerically solving the circuit model with the $\text{d}n_\text{QW}/\text{d}V_\text{FG}$ value at each $V_\text{FG}$ as an input, we can deduce $C_\text{SL}$ as a function of $V_\text{FG}$, as shown in Fig.~\ref{Fig3}(a). $C_\text{SL}$ decreases with decreasing $V_\text{FG}$, reflecting the disorder-broadened tail of the DOS.
Interestingly, $C_\text{SL}$ is not constant even at $V_\text{FG} > 0$ V, where it keeps increasing with $V_\text{FG}$.
Since quantum capacitance is proportional to the DOS at the Fermi level, the obtained $C_\text{SL}$ can be translated into the effective mass $m^\star_\text{SL}$ that would produce the same DOS for parabolic dispersion through the relation $D_\text{SL} = g_\text{SL}m^\star_\text{SL}$/$2\pi\hbar^2$.
In thin AlAs layers, quantum confinement and strain split the three-fold valley degeneracy in bulk into one and two, with the former becoming lower in energy for a thickness below $5.5$--$6.0$ nm \cite{Khisameeva2019,VanKesteren1989}.
We therefore assumed $g_\text{SL} = 2$, taking into account the spin degeneracy. The effective mass $m^\star_\text{SL}$ evaluated in this way is shown on the right axis of Fig.~\ref{Fig3}(a).
In bulk AlAs, the effective masses in the transverse and longitudinal directions of the ellipsoid Fermi surface are $0.22m_0$ and $0.97m_0$, respectively \cite{Vurgaftman2001}.
For AlAs QWs thinner than 6.0 nm, the 2DES occupies the lower non-degenerate valley, where experiments report a transverse mass of ($0.2$--$0.3$)$m_0$ \cite{Yamada1994,Momose1999,Vakili2004}.
The obtained $m^\star_\text{SL}$/$m_0$, which approaches the expected value ($0.2$--$0.3$) with increasing $V_\text{FG}$, is reasonable.

Once $C_\text{SL}$ is obtained as a function of $V_\text{FG}$, one can calculate the electron density in the AlAs layers [$n_\text{AlAs1(2)}$] and SL ($n_\text{SL} = n_\text{AlAs1} + n_\text{AlAs2}$). Figure~\ref{Fig3}(b) shows the $V_\text{FG}$ dependence of $n_\text{SL}$. The right axis of the figure indicates the excess electron density normalized by the doping density ($N_\text{Si} = 10^{16}$~m$^{-2}$).
While $n_\text{SL}$ varies almost linearly with $V_\text{FG}$, the slope decreases slightly at $V_\text{FG} < -1.0$~V.

\begin{figure}[ptb]
\includegraphics[scale=1.05]{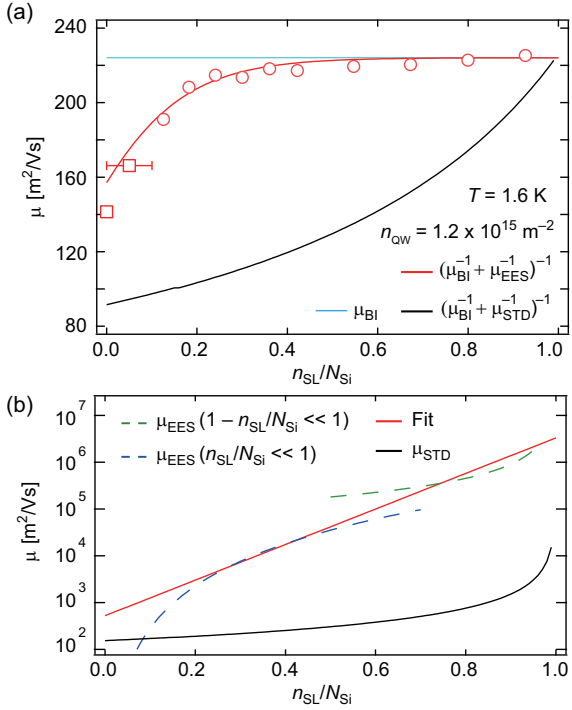}\caption{
(a) Mobility vs. $n_\text{SL}/N_\text{Si}$.
Open circles (squares) are experimental results for $n_\text{QW} = 1.2 \times 10^{15}$m$^{-2}$ measured at 1.6 K with $V_\text{FG}$ set at 4.3 K (room temperature). The error bar represents the uncertainty in $n_\text{SL}/N_\text{Si}$ for the data taken at $V_\text{FG} = -1.70$~V.
The red line is the total mobility calculated using RI-limited mobility with excess-electron screening ($\mu_\text{EES}$) shown in (b) and fitted to experimental data using BI-limited mobility ($\mu_\text{BI}$) as a parameter.
The value of $\mu_\text{BI}$ used for the fit is shown by the cyan line.
The black line shows the total mobility for the standard model without excess electron screening ($\mu_\text{STD}$) shown in (b).
(b) RI-limited mobility calculated with various models. The dashed green and blue curves were respectively calculated using the analytical formula of $\mu_\text{EES}$ for the two cases, $n_\text{SL}/N_\text{Si} \ll 1$ and ($1 - n_\text{SL}/N_\text{Si}) \ll 1$ \cite{SammonChe2018}. The solid red curve is a fit to the two regimes using an empirical formula (see main text for details). The black curve was calculated using the standard model.
}
\label{Fig4}
\end{figure}

\subsection{C. Effects on mobility}
Now let us investigate the effect of screening on mobility.
The symbols in Fig.~\ref{Fig4}(a) show the 1.6-K mobility ($\mu$) measured at the same carrier density ($n_\text{QW} = 1.2 \times 10^{15}$ m$^{-2}$) but with the sample prepared to have different $n_\text{SL}/N_\text{Si}$ values.
The open circles show data obtained by re-adjusting $n_\text{QW}$ with the back gate after equilibrating the system at 4.3~K for $V_\text{FG}$ ($-1.3$--$0$~V) and cooling the sample to 1.6~K.
For this $V_\text{FG}$ range, $n_\text{SL}/N_\text{Si}$ varied between 0.13 and 0.93 [see Fig.~\ref{Fig3}(b)]. As equilibration was not obtained for $V_\text{FG} < -1.3$~V at 4.3~K, we employed a different method to achieve smaller $n_\text{SL}/N_\text{Si}$. The open squares in Fig.~\ref{Fig4}(a) show data obtained by applying $V_\text{FG} \leq -1.7$~V at room temperature and re-adjusting $n_\text{QW}$ at 1.6~K with the back gate.
We confirmed that the SL was already depleted (i.e., $n_\text{SL}/N_\text{Si} = 0$) for $V_\text{FG} = -1.88$~V by noting that at 1.6~K the 2DES was depleted at zero back gate voltage. As Fig.~\ref{Fig4}(a) shows, $\mu$ decreased by 37\% as $n_\text{SL}/N_\text{Si}$ decreased from 0.93 to 0.

We characterize the $n_\text{SL}/N_\text{Si}$ dependence by considering two main sources of disorder in modulation-doped GaAs 2DESs, i.e., background ionized impurities (BIs) and remote ionized impurities (RIs).
For the mobility limited by RIs, we used the excess electron screening (EES) model proposed by Sammon \textit{et al.}~\cite{SammonChe2018}.
The dashed blue and green curves in Fig.~\ref{Fig4}(b) show the mobility $\mu_\text{EES}$ calculated using the EES model for the two limits, $n_\text{SL}/N_\text{Si} \ll 1$ and $1 - n_\text{SL}/N_\text{Si} \ll 1$, respectively.
The red curve is the fitting using the empirical formula derived in Ref.~\cite{SammonChe2018}. By adjusting the numerical parameters to connect the two limits, we obtain
\begin{equation}
	\mu_{\text{EES}} = \frac{e}{\hbar}k_\text{F}^3d^5\times10^{3.5\frac{n_{\text{SL}}}{N_{\text{Si}}}-1.25}
\end{equation}
Here, $k_\text{F} = (2\pi n_\text{QW})^{1/2}$ is the Fermi wave number and $d = 90$ nm is the center-to-center distance between the QW and SL.
The red curve in Fig.~\ref{Fig4}(a) shows the least-squares fit of the total mobility $(1$/$\mu_\text{BI} + 1$/$\mu_\text{EES})^{-1}$ based on Matthiessen's rule, where $\mu_\text{BI}$ is the only parameter and we assumed it to be constant.
The $\mu_\text{BI}$ value obtained from the fit is shown by the cyan line in Fig.~\ref{Fig4}(a).
The EES model well explains the overall $n_\text{SL}/N_\text{Si}$ dependence of the measured $\mu$, providing good agreement for $n_\text{SL}/N_\text{Si} \geq 0.13$. For $n_\text{SL}/N_\text{Si} < 0.13$, the agreement between the experiment and model becomes less satisfactory, which is because we tried to fit both regimes of $n_\text{SL}/N_\text{Si} \ll 1$ and $1 - n_\text{SL}/N_\text{Si} \ll 1$ using Eq.~(1) [Fig.~\ref{Fig4}(b)].
For comparison, we also calculated the mobility using the standard model ($\mu_\text{STD}$) \cite{Hirakawa1986}, assuming independent scattering by ($N_\text{Si} - n_\text{SL}$) ionized donors.
The black lines in Figs.~\ref{Fig4}(b) and ~\ref{Fig4}(a) show $\mu_\text{STD}$ and the resultant total mobility $(1/\mu_\text{BI} + 1/\mu_\text{STD})^{-1}$, respectively.
The independent-scattering model predicts a mobility way below the experimental result, which in turn demonstrates the importance of the screening by excess electrons.

We also examined the possibility of excess electrons in the SL affecting $\mu_\text{BI}$. Sammon \textit{et al.} reported that, for strong screening, the contribution of BIs to $\mu_\text{BI}$ is canceled out by the image-charge effect when they are located farther than $0.5d$ from the center of the QW \cite{SammonZu2018}. We calculated $\mu_\text{BI}$ by integrating contributions from BIs over different spatial ranges, $0.5d$ and $d$. The difference between the two cases is less than 1\%, thus corroborating our assumption of constant $\mu_\text{BI}$.

\subsection{D. Quantum lifetime}
\begin{figure}[ptb]
\includegraphics[scale=1.05]{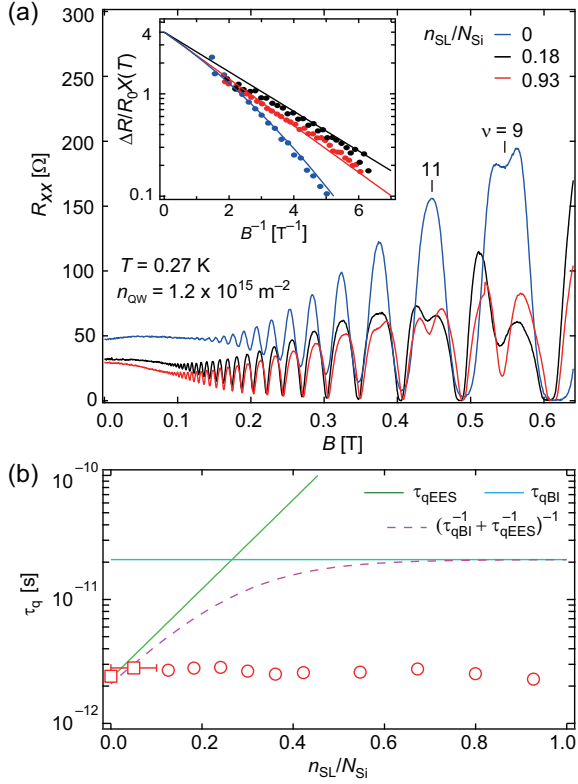}\caption{
(a) Shubnikov-de Haas oscillations under strong (red), intermediate (black), and weak (blue) screening conditions for the same density. The inset shows the Dingle plot for each case. (b) Quantum lifetime vs. $n_\text{SL}/N_\text{Si}$. Open circles (squares) are experimental results measured at 0.27 K with $V_\text{FG}$ set at 4.3 K (room temperature). The error bar represents the uncertainty in $n_\text{SL}/N_\text{Si}$ for the data taken at $V_\text{FG} = -1.70$~V. The cyan and green lines are calculated lifetimes limited by BIs ($\tau_\text{qBI}$) and RIs ($\tau_\text{qEES}$). The dashed magenta line shows the total quantum lifetime.
}
\label{Fig5}
\end{figure}

Finally, let us investigate the effect of screening on quantum lifetime ($\tau_\text{q}$) deduced from Shubnikov–de Haas (SdH) oscillations, a quantity often argued to be a better indicator of sample quality than mobility in terms of FQHEs.
Figure~\ref{Fig5}(a) shows the SdH oscillations measured at 0.27 K at a constant carrier density ($n_\text{QW} = 1.2 \times 10^{15}$~m$^{-2}$) under different screening conditions of $n_\text{SL}/N_\text{Si} = 0.93$, $0.18$, and $0$ (corresponding $V_\text{FG}$ of 0, $-1.20$, and $-1.88$~V, respectively).
Under the well-screened condition ($n_\text{SL}/N_\text{Si} = 0.93$), minima at odd filling factors due to spin splitting are more pronounced, indicating the influence of screening.
We extracted the quantum lifetime $\tau_\text{q}$ by using the functional form of SdH oscillations, given as \cite{Coleridge1991}
\begin{equation}
	\Delta R = 4R_0\text{exp}\left(\frac{-\pi}{\omega_\text{c}\tau_\text{q}}\right)\chi(T),\\
\end{equation}
with
\begin{equation}
	\chi(T) = \cfrac{\cfrac{2\pi^2k_\text{B}T}{\hbar\omega_\text{c}}}{\text{sinh}{\cfrac{2\pi^2k_\text{B}T}{\hbar\omega_\text{c}}}}.\\
\end{equation}
Here, $\Delta R$ is the amplitude of the SdH oscillations, $\omega_\text{c}$ is the cyclotron frequency, $R_0$  is the $R_{xx}$ at zero magnetic field, $\chi(T)$ is a thermal damping factor, and $k_\text{B}$ is the Boltzmann constant.
Thus, the slope of $\Delta R$/$R_0\chi(T)$ vs. $1/B$, known as a Dingle plot, shown in the inset of Fig.~\ref{Fig5}(a) gives the quantum lifetime $\tau_\text{q}$
\footnote{
For the data in the weak screening regime taken with $V_\text{FG} \leq -1.7$~V set at room temperature, analysis taking into account density inhomogeneity \cite{Qian2017} was necessary to fit the Dingle plot with the correct intercept of 4 at $B^{-1}=0$. The density inhomogeneity derived from the fit was 0.6 and 1.8\% for $V_\text{FG}=-1.70$ and $-1.88$~V, respectively.
}.
The obtained $\tau_\text{q}$ is shown by open symbols in Fig.~\ref{Fig5}(b).
In contrast to $\mu$, or transport lifetime ($\tau_\text{t} = m_\text{QW}\mu/e$), $\tau_\text{q}$ does not show a discernible change as a function of $n_\text{SL}/N_\text{Si}$.

We compared the measured $\tau_\text{q}$ with the calculated quantum lifetimes limited by RIs and BIs ($\tau_\text{qEES}$ and $\tau_\text{qBI}$), which are shown in Fig.~\ref{Fig5}(b) by the green and cyan lines, respectively. Here, $\tau_\text{qEES}$ was calculated with the EES model \cite{SammonChe2018}, whereas $\tau_\text{qBI}$ was calculated with the independent-scattering model using the BI concentration obtained from $\mu_\text{BI}$ ($= 224$~m$^{2}$/Vs).
The expected total quantum lifetime is shown by the dashed magenta line. Although the measured $\tau_\text{q}$ is close to the values expected in the weak screening regime, it remains about $2.6$~ps when $n_\text{SL}/N_\text{Si}$ increases, much lower than expected in the intermediate and strong screening regimes. This discrepancy was not mitigated even when a more elaborate model was employed, such as one with different BI concentrations in the GaAs QW and AlGaAs barrier layers \cite{SammonZu2018} and remote charges on the sample surface and the back gate \cite{Chen2012,Wang2013}.
Extrinsic mechanisms that might reduce the apparent quantum lifetime, such as the finite density gradient \cite{Qian2017} \footnote{An attempt to fit the Dingle-plot data in Fig.~\ref{Fig5}(a) using the density-gradient model in Ref. \cite{Qian2017} together with the calculated quantum lifetime resulted in a strongly nonlinear curve, which did not fit the experimental results.} and the response time of the lock-in amplifier \footnote{Reducing the field sweep rate from 10 mT/sec (which we normally use) to 10 $\mu$T/sec did not affect the measured value of $\mu_\text{q}$.}, did not explain the discrepancy, either.
Ultra-high-quality samples with much longer $\tau_\text{q}$, such as those in Refs.~\cite{Qian2017,Fu2018}, might be necessary to observe the predicted screening effect on $\tau_\text{q}$.
Yet, it is interesting that the visibility of the spin gap varies with $n_\text{SL}/N_\text{Si}$ even when $\tau_\text{q}$ remains constant, as we observed.
For $n_\text{SL}/N_\text{Si} = 0$, the density inhomogeneity estimated from the analysis of the Dingle plot \cite{Qian2017} is 1.8\%, which may be partly responsible for the poorly developed quantum Hall effects at even as well as odd integer fillings. However, for $n_\text{SL}/N_\text{Si} = 0.18$ and $0.93$, the estimated density inhomogeneity is less than 0.2\% with no clear difference, which cannot account for the difference in the visibility of the spin gap.
This suggests that the screening of long-range disorder becomes more important for interaction phenomena such as FQHEs.
The broad $R_{xx}$ minimum around $B = 8$ T seen in Fig.~\ref{Fig1} is a precursor of the $\nu = 2/3$ FQHE.
By investigating FQHEs at lower temperatures under different screening conditions, it will be possible to examine how the energy gap of FQHEs is correlated with the composite fermion mobility \cite{Kang1995} deduced from the resistivity at $\nu = 1/2$.

\section{III. CONCLUSION}
In summary, we investigated the screening effects of SL doping on the mobility and quantum lifetime of a GaAs 2DES by controlling the excess electron density in the SL with a top gate.
The dependence of mobility on excess electron density is consistent with theory taking into account the screening effect.
On the other hand, the measured quantum mobility was much lower than expected from theory and did not show a discernible change with excess electron density.
The excess electrons also affected the depth of the spin-gap minima in the Shubnikov-de Haas oscillations, which suggests the possibility of controlling the visibility of FQHEs in-situ.

\section{ACKNOWLEDGEMENTS}
The authors thank M. Kamiya and H. Irie for support in the measurements, H. Murofushi for processing the device. and M. A. Zudov for helpful discussions. This work was supported by a JSPS KAKENHI Grant, No. JP15H05854.


\begin{thebibliography}{37}%
\makeatletter
\providecommand \@ifxundefined [1]{%
 \@ifx{#1\undefined}
}%
\providecommand \@ifnum [1]{%
 \ifnum #1\expandafter \@firstoftwo
 \else \expandafter \@secondoftwo
 \fi
}%
\providecommand \@ifx [1]{%
 \ifx #1\expandafter \@firstoftwo
 \else \expandafter \@secondoftwo
 \fi
}%
\providecommand \natexlab [1]{#1}%
\providecommand \enquote  [1]{``#1''}%
\providecommand \bibnamefont  [1]{#1}%
\providecommand \bibfnamefont [1]{#1}%
\providecommand \citenamefont [1]{#1}%
\providecommand \href@noop [0]{\@secondoftwo}%
\providecommand \href [0]{\begingroup \@sanitize@url \@href}%
\providecommand \@href[1]{\@@startlink{#1}\@@href}%
\providecommand \@@href[1]{\endgroup#1\@@endlink}%
\providecommand \@sanitize@url [0]{\catcode `\\12\catcode `\$12\catcode
  `\&12\catcode `\#12\catcode `\^12\catcode `\_12\catcode `\%12\relax}%
\providecommand \@@startlink[1]{}%
\providecommand \@@endlink[0]{}%
\providecommand \url  [0]{\begingroup\@sanitize@url \@url }%
\providecommand \@url [1]{\endgroup\@href {#1}{\urlprefix }}%
\providecommand \urlprefix  [0]{URL }%
\providecommand \Eprint [0]{\href }%
\providecommand \doibase [0]{http://dx.doi.org/}%
\providecommand \selectlanguage [0]{\@gobble}%
\providecommand \bibinfo  [0]{\@secondoftwo}%
\providecommand \bibfield  [0]{\@secondoftwo}%
\providecommand \translation [1]{[#1]}%
\providecommand \BibitemOpen [0]{}%
\providecommand \bibitemStop [0]{}%
\providecommand \bibitemNoStop [0]{.\EOS\space}%
\providecommand \EOS [0]{\spacefactor3000\relax}%
\providecommand \BibitemShut  [1]{\csname bibitem#1\endcsname}%
\let\auto@bib@innerbib\@empty
\bibitem [{\citenamefont {Pfeiffer}\ and\ \citenamefont
  {West}(2003)}]{Pfeiffer2003}%
  \BibitemOpen
  \bibfield  {author} {\bibinfo {author} {\bibfnamefont {L.}~\bibnamefont
  {Pfeiffer}}\ and\ \bibinfo {author} {\bibfnamefont {K.~W.}\ \bibnamefont
  {West}},\ }\bibfield  {title} {\enquote {\bibinfo {title} {{The role of MBE
  in recent quantum Hall effect physics discoveries}},}\ }\href {\doibase
  10.1016/j.physe.2003.09.035} {\bibfield  {journal} {\bibinfo  {journal}
  {Physica E}\ }\textbf {\bibinfo {volume} {20}},\ \bibinfo {pages} {57}
  (\bibinfo {year} {2003})}\BibitemShut {NoStop}%
\bibitem [{\citenamefont {Umansky}\ \emph {et~al.}(2009)\citenamefont
  {Umansky}, \citenamefont {Heiblum}, \citenamefont {Levinson}, \citenamefont
  {Smet}, \citenamefont {N{\"{u}}bler},\ and\ \citenamefont
  {Dolev}}]{Umansky2009}%
  \BibitemOpen
  \bibfield  {author} {\bibinfo {author} {\bibfnamefont {V.}~\bibnamefont
  {Umansky}}, \bibinfo {author} {\bibfnamefont {M.}~\bibnamefont {Heiblum}},
  \bibinfo {author} {\bibfnamefont {Y.}~\bibnamefont {Levinson}}, \bibinfo
  {author} {\bibfnamefont {J.}~\bibnamefont {Smet}}, \bibinfo {author}
  {\bibfnamefont {J.}~\bibnamefont {N{\"{u}}bler}}, \ and\ \bibinfo {author}
  {\bibfnamefont {M.}~\bibnamefont {Dolev}},\ }\bibfield  {title} {\enquote
  {\bibinfo {title} {{MBE growth of ultra-low disorder 2DEG with mobility
  exceeding $35 \times 10^{6}~\text{cm}^{2}\text{/Vs}$}},}\ }\href {\doibase
  10.1016/j.jcrysgro.2008.09.151} {\bibfield  {journal} {\bibinfo  {journal}
  {J. Cryst. Growth}\ }\textbf {\bibinfo {volume} {311}},\ \bibinfo {pages}
  {1658} (\bibinfo {year} {2009})}\BibitemShut {NoStop}%
\bibitem [{\citenamefont {Gardner}\ \emph {et~al.}(2016)\citenamefont
  {Gardner}, \citenamefont {Fallahi}, \citenamefont {Watson},\ and\
  \citenamefont {Manfra}}]{Gardner2016}%
  \BibitemOpen
  \bibfield  {author} {\bibinfo {author} {\bibfnamefont {G.~C.}\ \bibnamefont
  {Gardner}}, \bibinfo {author} {\bibfnamefont {S.}~\bibnamefont {Fallahi}},
  \bibinfo {author} {\bibfnamefont {J.~D.}\ \bibnamefont {Watson}}, \ and\
  \bibinfo {author} {\bibfnamefont {M.~J.}\ \bibnamefont {Manfra}},\ }\bibfield
   {title} {\enquote {\bibinfo {title} {{Modified MBE hardware and techniques
  and role of gallium purity for attainment of two dimensional electron gas
  mobility $> 35 \times 10^{6}~\text{cm}^{2}\text{/Vs}$ in AlGaAs/GaAs quantum
  wells grown by MBE}},}\ }\href {\doibase 10.1016/j.jcrysgro.2016.02.010}
  {\bibfield  {journal} {\bibinfo  {journal} {J. Cryst. Growth}\ }\textbf
  {\bibinfo {volume} {441}},\ \bibinfo {pages} {71} (\bibinfo {year}
  {2016})}\BibitemShut {NoStop}%
\bibitem [{\citenamefont {Manfra}(2014)}]{Manfra2014}%
  \BibitemOpen
  \bibfield  {author} {\bibinfo {author} {\bibfnamefont {M.~J.}\ \bibnamefont
  {Manfra}},\ }\bibfield  {title} {\enquote {\bibinfo {title} {{Molecular Beam
  Epitaxy of Ultra-High-Quality AlGaAs/GaAs Heterostructures: Enabling Physics
  in Low-Dimensional Electronic Systems}},}\ }\href {\doibase
  10.1146/annurev-conmatphys-031113-133905} {\bibfield  {journal} {\bibinfo
  {journal} {Annu. Rev. Condens. Matter Phys.}\ }\textbf {\bibinfo {volume}
  {5}},\ \bibinfo {pages} {347} (\bibinfo {year} {2014})}\BibitemShut {NoStop}%
\bibitem [{\citenamefont {Chung}\ \emph {et~al.}(2020)\citenamefont {Chung},
  \citenamefont {Rosales}, \citenamefont {Baldwin}, \citenamefont {West},
  \citenamefont {Shayegan},\ and\ \citenamefont {Pfeiffer}}]{Chung2020}%
  \BibitemOpen
  \bibfield  {author} {\bibinfo {author} {\bibfnamefont {Y.~J.}\ \bibnamefont
  {Chung}}, \bibinfo {author} {\bibfnamefont {K.~A.~Villegas}\ \bibnamefont
  {Rosales}}, \bibinfo {author} {\bibfnamefont {K.~W.}\ \bibnamefont
  {Baldwin}}, \bibinfo {author} {\bibfnamefont {K.~W.}\ \bibnamefont {West}},
  \bibinfo {author} {\bibfnamefont {M.}~\bibnamefont {Shayegan}}, \ and\
  \bibinfo {author} {\bibfnamefont {L.~N.}\ \bibnamefont {Pfeiffer}},\
  }\bibfield  {title} {\enquote {\bibinfo {title} {{Working principles of
  doping-well structures for high-mobility two-dimensional electron
  systems}},}\ }\href {\doibase 10.1103/PhysRevMaterials.4.044003} {\bibfield
  {journal} {\bibinfo  {journal} {Phys. Rev. Mater.}\ }\textbf {\bibinfo
  {volume} {4}},\ \bibinfo {pages} {44003} (\bibinfo {year}
  {2020})}\BibitemShut {NoStop}%
\bibitem [{\citenamefont {Willett}\ \emph {et~al.}(1988)\citenamefont
  {Willett}, \citenamefont {Stormer}, \citenamefont {Tsui}, \citenamefont
  {Pfeiffer}, \citenamefont {West},\ and\ \citenamefont
  {Baldwin}}]{RLWillett1988}%
  \BibitemOpen
  \bibfield  {author} {\bibinfo {author} {\bibfnamefont {R.~L.}\ \bibnamefont
  {Willett}}, \bibinfo {author} {\bibfnamefont {H.~L.}\ \bibnamefont
  {Stormer}}, \bibinfo {author} {\bibfnamefont {D.~C.}\ \bibnamefont {Tsui}},
  \bibinfo {author} {\bibfnamefont {L.~N.}\ \bibnamefont {Pfeiffer}}, \bibinfo
  {author} {\bibfnamefont {K.~W.}\ \bibnamefont {West}}, \ and\ \bibinfo
  {author} {\bibfnamefont {K.~W.}\ \bibnamefont {Baldwin}},\ }\bibfield
  {title} {\enquote {\bibinfo {title} {Termination of the series of fractional
  quantum hall states at small filling factors},}\ }\href {\doibase
  10.1103/PhysRevB.38.7881} {\bibfield  {journal} {\bibinfo  {journal} {Phys.
  Rev. B}\ }\textbf {\bibinfo {volume} {38}},\ \bibinfo {pages} {7881}
  (\bibinfo {year} {1988})}\BibitemShut {NoStop}%
\bibitem [{\citenamefont {Goldman}\ \emph {et~al.}(1990)\citenamefont
  {Goldman}, \citenamefont {Santos}, \citenamefont {Shayegan},\ and\
  \citenamefont {Cunningham}}]{Goldman1990}%
  \BibitemOpen
  \bibfield  {author} {\bibinfo {author} {\bibfnamefont {V.~J.}\ \bibnamefont
  {Goldman}}, \bibinfo {author} {\bibfnamefont {M.}~\bibnamefont {Santos}},
  \bibinfo {author} {\bibfnamefont {M.}~\bibnamefont {Shayegan}}, \ and\
  \bibinfo {author} {\bibfnamefont {J.~E.}\ \bibnamefont {Cunningham}},\
  }\bibfield  {title} {\enquote {\bibinfo {title} {{Evidence for
  two-dimentional quantum Wigner crystal}},}\ }\href {\doibase
  10.1103/PhysRevLett.65.2189} {\bibfield  {journal} {\bibinfo  {journal}
  {Phys. Rev. Lett.}\ }\textbf {\bibinfo {volume} {65}},\ \bibinfo {pages}
  {2189} (\bibinfo {year} {1990})}\BibitemShut {NoStop}%
\bibitem [{\citenamefont {Lilly}\ \emph {et~al.}(1999)\citenamefont {Lilly},
  \citenamefont {Cooper}, \citenamefont {Eisenstein}, \citenamefont
  {Pfeiffer},\ and\ \citenamefont {West}}]{Lilly1999}%
  \BibitemOpen
  \bibfield  {author} {\bibinfo {author} {\bibfnamefont {M.~P.}\ \bibnamefont
  {Lilly}}, \bibinfo {author} {\bibfnamefont {K.~B.}\ \bibnamefont {Cooper}},
  \bibinfo {author} {\bibfnamefont {J.~P.}\ \bibnamefont {Eisenstein}},
  \bibinfo {author} {\bibfnamefont {L.~N.}\ \bibnamefont {Pfeiffer}}, \ and\
  \bibinfo {author} {\bibfnamefont {K.~W.}\ \bibnamefont {West}},\ }\bibfield
  {title} {\enquote {\bibinfo {title} {{Evidence for an anisotropic state of
  two-dimensional electrons in high landau levels}},}\ }\href {\doibase
  10.1103/PhysRevLett.82.394} {\bibfield  {journal} {\bibinfo  {journal} {Phys.
  Rev. Lett.}\ }\textbf {\bibinfo {volume} {82}},\ \bibinfo {pages} {394}
  (\bibinfo {year} {1999})}\BibitemShut {NoStop}%
\bibitem [{\citenamefont {Du}\ \emph {et~al.}(1999)\citenamefont {Du},
  \citenamefont {Tsui}, \citenamefont {Stormer}, \citenamefont {Pfeiffer},
  \citenamefont {Baldwin},\ and\ \citenamefont {West}}]{Du1999}%
  \BibitemOpen
  \bibfield  {author} {\bibinfo {author} {\bibfnamefont {R.~R.}\ \bibnamefont
  {Du}}, \bibinfo {author} {\bibfnamefont {D.~C.}\ \bibnamefont {Tsui}},
  \bibinfo {author} {\bibfnamefont {H.~L.}\ \bibnamefont {Stormer}}, \bibinfo
  {author} {\bibfnamefont {L.~N.}\ \bibnamefont {Pfeiffer}}, \bibinfo {author}
  {\bibfnamefont {K.~W.}\ \bibnamefont {Baldwin}}, \ and\ \bibinfo {author}
  {\bibfnamefont {K.~W.}\ \bibnamefont {West}},\ }\bibfield  {title} {\enquote
  {\bibinfo {title} {Strongly anisotropic transport in higher two-dimensional
  landau levels},}\ }\href {\doibase
  https://doi.org/10.1016/S0038-1098(98)00578-X} {\bibfield  {journal}
  {\bibinfo  {journal} {Solid State Commun.}\ }\textbf {\bibinfo {volume}
  {109}},\ \bibinfo {pages} {389} (\bibinfo {year} {1999})}\BibitemShut
  {NoStop}%
\bibitem [{\citenamefont {Suen}\ \emph {et~al.}(1992)\citenamefont {Suen},
  \citenamefont {Engel}, \citenamefont {Santos}, \citenamefont {Shayegan},\
  and\ \citenamefont {Tsui}}]{Suen1992}%
  \BibitemOpen
  \bibfield  {author} {\bibinfo {author} {\bibfnamefont {Y.~W.}\ \bibnamefont
  {Suen}}, \bibinfo {author} {\bibfnamefont {L.~W.}\ \bibnamefont {Engel}},
  \bibinfo {author} {\bibfnamefont {M.~B.}\ \bibnamefont {Santos}}, \bibinfo
  {author} {\bibfnamefont {M.}~\bibnamefont {Shayegan}}, \ and\ \bibinfo
  {author} {\bibfnamefont {D.~C.}\ \bibnamefont {Tsui}},\ }\bibfield  {title}
  {\enquote {\bibinfo {title} {{Observation of a $\nu$ = 1/2 fractional quantum
  Hall state in a double-layer electron system}},}\ }\href {\doibase
  10.1103/PhysRevLett.68.1379} {\bibfield  {journal} {\bibinfo  {journal}
  {Phys. Rev. Lett.}\ }\textbf {\bibinfo {volume} {68}},\ \bibinfo {pages}
  {1379} (\bibinfo {year} {1992})}\BibitemShut {NoStop}%
\bibitem [{\citenamefont {Tsui}\ \emph {et~al.}(1982)\citenamefont {Tsui},
  \citenamefont {Stormer},\ and\ \citenamefont {Gossard}}]{Tsui1982}%
  \BibitemOpen
  \bibfield  {author} {\bibinfo {author} {\bibfnamefont {D.~C.}\ \bibnamefont
  {Tsui}}, \bibinfo {author} {\bibfnamefont {H.~L.}\ \bibnamefont {Stormer}}, \
  and\ \bibinfo {author} {\bibfnamefont {A.~C.}\ \bibnamefont {Gossard}},\
  }\bibfield  {title} {\enquote {\bibinfo {title} {Two-dimensional
  magnetotransport in the extreme quantum limit},}\ }\href {\doibase
  10.1103/PhysRevLett.48.1559} {\bibfield  {journal} {\bibinfo  {journal}
  {Phys. Rev. Lett.}\ }\textbf {\bibinfo {volume} {48}},\ \bibinfo {pages}
  {1559} (\bibinfo {year} {1982})}\BibitemShut {NoStop}%
\bibitem [{\citenamefont {Willett}\ \emph {et~al.}(1987)\citenamefont
  {Willett}, \citenamefont {Eisenstein}, \citenamefont {St{\"{o}}rmer},
  \citenamefont {Tsui}, \citenamefont {Gossard},\ and\ \citenamefont
  {English}}]{Willett1987}%
  \BibitemOpen
  \bibfield  {author} {\bibinfo {author} {\bibfnamefont {R.}~\bibnamefont
  {Willett}}, \bibinfo {author} {\bibfnamefont {J.~P.}\ \bibnamefont
  {Eisenstein}}, \bibinfo {author} {\bibfnamefont {H.~L.}\ \bibnamefont
  {St{\"{o}}rmer}}, \bibinfo {author} {\bibfnamefont {D.~C.}\ \bibnamefont
  {Tsui}}, \bibinfo {author} {\bibfnamefont {A.~C.}\ \bibnamefont {Gossard}}, \
  and\ \bibinfo {author} {\bibfnamefont {J.~H.}\ \bibnamefont {English}},\
  }\bibfield  {title} {\enquote {\bibinfo {title} {{Observation of an
  even-denominator quantum number in the fractional quantum Hall effect}},}\
  }\href {\doibase 10.1103/PhysRevLett.59.1776} {\bibfield  {journal} {\bibinfo
   {journal} {Phys. Rev. Lett.}\ }\textbf {\bibinfo {volume} {59}},\ \bibinfo
  {pages} {1776} (\bibinfo {year} {1987})}\BibitemShut {NoStop}%
\bibitem [{\citenamefont {Pan}\ \emph {et~al.}(2011)\citenamefont {Pan},
  \citenamefont {Masuhara}, \citenamefont {Sullivan}, \citenamefont {Baldwin},
  \citenamefont {West}, \citenamefont {Pfeiffer},\ and\ \citenamefont
  {Tsui}}]{Pan2011}%
  \BibitemOpen
  \bibfield  {author} {\bibinfo {author} {\bibfnamefont {W.}~\bibnamefont
  {Pan}}, \bibinfo {author} {\bibfnamefont {N.}~\bibnamefont {Masuhara}},
  \bibinfo {author} {\bibfnamefont {N.~S.}\ \bibnamefont {Sullivan}}, \bibinfo
  {author} {\bibfnamefont {K.~W.}\ \bibnamefont {Baldwin}}, \bibinfo {author}
  {\bibfnamefont {K.~W.}\ \bibnamefont {West}}, \bibinfo {author}
  {\bibfnamefont {L.~N.}\ \bibnamefont {Pfeiffer}}, \ and\ \bibinfo {author}
  {\bibfnamefont {D.~C.}\ \bibnamefont {Tsui}},\ }\bibfield  {title} {\enquote
  {\bibinfo {title} {{Impact of disorder on the 5/2 fractional quantum Hall
  state}},}\ }\href {\doibase 10.1103/PhysRevLett.106.206806} {\bibfield
  {journal} {\bibinfo  {journal} {Phys. Rev. Lett.}\ }\textbf {\bibinfo
  {volume} {106}},\ \bibinfo {pages} {206806} (\bibinfo {year}
  {2011})}\BibitemShut {NoStop}%
\bibitem [{\citenamefont {Gamez}\ and\ \citenamefont
  {Muraki}(2013)}]{Gamez2013}%
  \BibitemOpen
  \bibfield  {author} {\bibinfo {author} {\bibfnamefont {G.}~\bibnamefont
  {Gamez}}\ and\ \bibinfo {author} {\bibfnamefont {K.}~\bibnamefont {Muraki}},\
  }\bibfield  {title} {\enquote {\bibinfo {title} {{$\nu$ = 5/2 fractional
  quantum Hall state in low-mobility electron systems: Different roles of
  disorder}},}\ }\href {\doibase 10.1103/PhysRevB.88.075308} {\bibfield
  {journal} {\bibinfo  {journal} {Phys. Rev. B}\ }\textbf {\bibinfo {volume}
  {88}},\ \bibinfo {pages} {075308} (\bibinfo {year} {2013})}\BibitemShut
  {NoStop}%
\bibitem [{\citenamefont {Friedland}\ \emph {et~al.}(1996)\citenamefont
  {Friedland}, \citenamefont {Hey}, \citenamefont {Kostial}, \citenamefont
  {Klann},\ and\ \citenamefont {Ploog}}]{Friedland1996}%
  \BibitemOpen
  \bibfield  {author} {\bibinfo {author} {\bibfnamefont {K.~J.}\ \bibnamefont
  {Friedland}}, \bibinfo {author} {\bibfnamefont {R.}~\bibnamefont {Hey}},
  \bibinfo {author} {\bibfnamefont {H.}~\bibnamefont {Kostial}}, \bibinfo
  {author} {\bibfnamefont {R.}~\bibnamefont {Klann}}, \ and\ \bibinfo {author}
  {\bibfnamefont {K.}~\bibnamefont {Ploog}},\ }\bibfield  {title} {\enquote
  {\bibinfo {title} {{New concept for the reduction of impurity scattering in
  remotely doped GaAs quantum wells}},}\ }\href {\doibase
  10.1103/PhysRevLett.77.4616} {\bibfield  {journal} {\bibinfo  {journal}
  {Phys. Rev. Lett.}\ }\textbf {\bibinfo {volume} {77}},\ \bibinfo {pages}
  {4616} (\bibinfo {year} {1996})}\BibitemShut {NoStop}%
\bibitem [{\citenamefont {Sammon}\ \emph
  {et~al.}(2018{\natexlab{a}})\citenamefont {Sammon}, \citenamefont {Zudov},\
  and\ \citenamefont {Shklovskii}}]{SammonChe2018}%
  \BibitemOpen
  \bibfield  {author} {\bibinfo {author} {\bibfnamefont {M.}~\bibnamefont
  {Sammon}}, \bibinfo {author} {\bibfnamefont {M.~A.}\ \bibnamefont {Zudov}}, \
  and\ \bibinfo {author} {\bibfnamefont {B.~I.}\ \bibnamefont {Shklovskii}},\
  }\bibfield  {title} {\enquote {\bibinfo {title} {{Mobility and quantum
  mobility of modern GaAs/AlGaAs heterostructures}},}\ }\href {\doibase
  10.1103/PhysRevMaterials.2.064604} {\bibfield  {journal} {\bibinfo  {journal}
  {Phys. Rev. Mater.}\ }\textbf {\bibinfo {volume} {2}},\ \bibinfo {pages}
  {104001} (\bibinfo {year} {2018}{\natexlab{a}})}\BibitemShut {NoStop}%
\bibitem [{\citenamefont {Sammon}\ \emph
  {et~al.}(2018{\natexlab{b}})\citenamefont {Sammon}, \citenamefont {Zudov},\
  and\ \citenamefont {Shklovskii}}]{SammonZu2018}%
  \BibitemOpen
  \bibfield  {author} {\bibinfo {author} {\bibfnamefont {M.}~\bibnamefont
  {Sammon}}, \bibinfo {author} {\bibfnamefont {M.~A.}\ \bibnamefont {Zudov}}, \
  and\ \bibinfo {author} {\bibfnamefont {B.~I.}\ \bibnamefont {Shklovskii}},\
  }\bibfield  {title} {\enquote {\bibinfo {title} {{Mobility and quantum
  mobility of modern GaAs/AlGaAs heterostructures}},}\ }\href {\doibase
  10.1103/PhysRevMaterials.2.064604} {\bibfield  {journal} {\bibinfo  {journal}
  {Phys. Rev. Mater.}\ }\textbf {\bibinfo {volume} {2}},\ \bibinfo {pages}
  {064604} (\bibinfo {year} {2018}{\natexlab{b}})}\BibitemShut {NoStop}%
\bibitem [{\citenamefont {R{\"{o}}ssler}\ \emph {et~al.}(2010)\citenamefont
  {R{\"{o}}ssler}, \citenamefont {Feil}, \citenamefont {Mensch}, \citenamefont
  {Ihn}, \citenamefont {Ensslin}, \citenamefont {Schuh},\ and\ \citenamefont
  {Wegscheider}}]{Rossler2010}%
  \BibitemOpen
  \bibfield  {author} {\bibinfo {author} {\bibfnamefont {C.}~\bibnamefont
  {R{\"{o}}ssler}}, \bibinfo {author} {\bibfnamefont {T.}~\bibnamefont {Feil}},
  \bibinfo {author} {\bibfnamefont {P.}~\bibnamefont {Mensch}}, \bibinfo
  {author} {\bibfnamefont {T.}~\bibnamefont {Ihn}}, \bibinfo {author}
  {\bibfnamefont {K.}~\bibnamefont {Ensslin}}, \bibinfo {author} {\bibfnamefont
  {D.}~\bibnamefont {Schuh}}, \ and\ \bibinfo {author} {\bibfnamefont
  {W.}~\bibnamefont {Wegscheider}},\ }\bibfield  {title} {\enquote {\bibinfo
  {title} {{Gating of high-mobility two-dimensional electron gases in
  GaAs/AlGaAs heterostructures}},}\ }\href {\doibase
  10.1088/1367-2630/12/4/043007} {\bibfield  {journal} {\bibinfo  {journal}
  {New J. Phys.}\ }\textbf {\bibinfo {volume} {12}},\ \bibinfo {pages} {043007}
  (\bibinfo {year} {2010})}\BibitemShut {NoStop}%
\bibitem [{\citenamefont {Dmitriev}\ \emph {et~al.}(2012)\citenamefont
  {Dmitriev}, \citenamefont {Strygin}, \citenamefont {Bykov}, \citenamefont
  {Dietrich},\ and\ \citenamefont {Vitkalov}}]{Dmitriev2012}%
  \BibitemOpen
  \bibfield  {author} {\bibinfo {author} {\bibfnamefont {D.~V.}\ \bibnamefont
  {Dmitriev}}, \bibinfo {author} {\bibfnamefont {I.~S.}\ \bibnamefont
  {Strygin}}, \bibinfo {author} {\bibfnamefont {A.~A.}\ \bibnamefont {Bykov}},
  \bibinfo {author} {\bibfnamefont {S.}~\bibnamefont {Dietrich}}, \ and\
  \bibinfo {author} {\bibfnamefont {S.~A.}\ \bibnamefont {Vitkalov}},\
  }\bibfield  {title} {\enquote {\bibinfo {title} {{Transport relaxation time
  and quantum lifetime in selectively doped GaAs/AlAs heterostructures}},}\
  }\href {\doibase 10.1134/S0021364012080048} {\bibfield  {journal} {\bibinfo
  {journal} {JETP Letters}\ }\textbf {\bibinfo {volume} {95}},\ \bibinfo
  {pages} {420} (\bibinfo {year} {2012})}\BibitemShut {NoStop}%
\bibitem [{\citenamefont {Peters}\ \emph {et~al.}(2017)\citenamefont {Peters},
  \citenamefont {Tiemann}, \citenamefont {Reichl}, \citenamefont {F{\"{a}}lt},
  \citenamefont {Dietsche},\ and\ \citenamefont {Wegscheider}}]{Peters2017}%
  \BibitemOpen
  \bibfield  {author} {\bibinfo {author} {\bibfnamefont {S.}~\bibnamefont
  {Peters}}, \bibinfo {author} {\bibfnamefont {L.}~\bibnamefont {Tiemann}},
  \bibinfo {author} {\bibfnamefont {C.}~\bibnamefont {Reichl}}, \bibinfo
  {author} {\bibfnamefont {S.}~\bibnamefont {F{\"{a}}lt}}, \bibinfo {author}
  {\bibfnamefont {W.}~\bibnamefont {Dietsche}}, \ and\ \bibinfo {author}
  {\bibfnamefont {W.}~\bibnamefont {Wegscheider}},\ }\bibfield  {title}
  {\enquote {\bibinfo {title} {{Improvement of the transport properties of a
  high-mobility electron system by intentional parallel conduction}},}\ }\href
  {\doibase 10.1063/1.4975055} {\bibfield  {journal} {\bibinfo  {journal}
  {Appl. Phys. Lett.}\ }\textbf {\bibinfo {volume} {110}},\ \bibinfo {pages}
  {042106} (\bibinfo {year} {2017})}\BibitemShut {NoStop}%
\bibitem [{\citenamefont {Eisenstein}\ \emph {et~al.}(1994)\citenamefont
  {Eisenstein}, \citenamefont {Pfeiffer},\ and\ \citenamefont
  {West}}]{Eisenstein1994}%
  \BibitemOpen
  \bibfield  {author} {\bibinfo {author} {\bibfnamefont {J.~P.}\ \bibnamefont
  {Eisenstein}}, \bibinfo {author} {\bibfnamefont {L.~N.}\ \bibnamefont
  {Pfeiffer}}, \ and\ \bibinfo {author} {\bibfnamefont {K.~W.}\ \bibnamefont
  {West}},\ }\bibfield  {title} {\enquote {\bibinfo {title} {{Compressibility
  of the two-dimensional electron gas: Measurements of the zero-field exchange
  energy and fractional quantum Hall gap}},}\ }\href {\doibase
  10.1103/PhysRevB.50.1760} {\bibfield  {journal} {\bibinfo  {journal} {Phys.
  Rev. B}\ }\textbf {\bibinfo {volume} {50}},\ \bibinfo {pages} {1760}
  (\bibinfo {year} {1994})}\BibitemShut {NoStop}%
\bibitem [{\citenamefont {Khisameeva}\ \emph {et~al.}(2019)\citenamefont
  {Khisameeva}, \citenamefont {Shchepetilnikov}, \citenamefont {Muravev},
  \citenamefont {Gubarev}, \citenamefont {Frolov}, \citenamefont {Nefyodov},
  \citenamefont {Kukushkin}, \citenamefont {Reichl}, \citenamefont {Dietsche},\
  and\ \citenamefont {Wegscheider}}]{Khisameeva2019}%
  \BibitemOpen
  \bibfield  {author} {\bibinfo {author} {\bibfnamefont {A.~R.}\ \bibnamefont
  {Khisameeva}}, \bibinfo {author} {\bibfnamefont {A.~V.}\ \bibnamefont
  {Shchepetilnikov}}, \bibinfo {author} {\bibfnamefont {V.~M.}\ \bibnamefont
  {Muravev}}, \bibinfo {author} {\bibfnamefont {S.~I.}\ \bibnamefont
  {Gubarev}}, \bibinfo {author} {\bibfnamefont {D.~D.}\ \bibnamefont {Frolov}},
  \bibinfo {author} {\bibfnamefont {Yu~A.}\ \bibnamefont {Nefyodov}}, \bibinfo
  {author} {\bibfnamefont {I.~V.}\ \bibnamefont {Kukushkin}}, \bibinfo {author}
  {\bibfnamefont {C.}~\bibnamefont {Reichl}}, \bibinfo {author} {\bibfnamefont
  {W.}~\bibnamefont {Dietsche}}, \ and\ \bibinfo {author} {\bibfnamefont
  {W.}~\bibnamefont {Wegscheider}},\ }\bibfield  {title} {\enquote {\bibinfo
  {title} {{Achieving balance of valley occupancy in narrow AlAs quantum
  wells}},}\ }\href {\doibase 10.1063/1.5079511} {\bibfield  {journal}
  {\bibinfo  {journal} {J. Appl. Phys.}\ }\textbf {\bibinfo {volume} {125}},\
  \bibinfo {pages} {154501} (\bibinfo {year} {2019})}\BibitemShut {NoStop}%
\bibitem [{\citenamefont {{Van Kesteren}}\ \emph {et~al.}(1989)\citenamefont
  {{Van Kesteren}}, \citenamefont {Cosman}, \citenamefont {Dawson},
  \citenamefont {Moore},\ and\ \citenamefont {Foxon}}]{VanKesteren1989}%
  \BibitemOpen
  \bibfield  {author} {\bibinfo {author} {\bibfnamefont {H.~W.}\ \bibnamefont
  {{Van Kesteren}}}, \bibinfo {author} {\bibfnamefont {E.~C.}\ \bibnamefont
  {Cosman}}, \bibinfo {author} {\bibfnamefont {P.}~\bibnamefont {Dawson}},
  \bibinfo {author} {\bibfnamefont {K.~J.}\ \bibnamefont {Moore}}, \ and\
  \bibinfo {author} {\bibfnamefont {C.~T.}\ \bibnamefont {Foxon}},\ }\bibfield
  {title} {\enquote {\bibinfo {title} {{Order of the X conduction-band valleys
  in type-II GaAs/AlAs quantum wells}},}\ }\href {\doibase
  10.1103/PhysRevB.39.13426} {\bibfield  {journal} {\bibinfo  {journal} {Phys.
  Rev. B}\ }\textbf {\bibinfo {volume} {39}},\ \bibinfo {pages} {13426}
  (\bibinfo {year} {1989})}\BibitemShut {NoStop}%
\bibitem [{\citenamefont {Vurgaftman}\ \emph {et~al.}(2001)\citenamefont
  {Vurgaftman}, \citenamefont {Meyer},\ and\ \citenamefont
  {Ram-Mohan}}]{Vurgaftman2001}%
  \BibitemOpen
  \bibfield  {author} {\bibinfo {author} {\bibfnamefont {I.}~\bibnamefont
  {Vurgaftman}}, \bibinfo {author} {\bibfnamefont {J.~R.}\ \bibnamefont
  {Meyer}}, \ and\ \bibinfo {author} {\bibfnamefont {L.~R.}\ \bibnamefont
  {Ram-Mohan}},\ }\bibfield  {title} {\enquote {\bibinfo {title} {{Band
  parameters for III-V compound semiconductors and their alloys}},}\ }\href
  {\doibase 10.1063/1.1368156} {\bibfield  {journal} {\bibinfo  {journal} {J.
  Appl. Phys.}\ }\textbf {\bibinfo {volume} {89}},\ \bibinfo {pages} {5815}
  (\bibinfo {year} {2001})}\BibitemShut {NoStop}%
\bibitem [{\citenamefont {Yamada}\ \emph {et~al.}(1994)\citenamefont {Yamada},
  \citenamefont {Maezawa}, \citenamefont {Yuen},\ and\ \citenamefont
  {Stradling}}]{Yamada1994}%
  \BibitemOpen
  \bibfield  {author} {\bibinfo {author} {\bibfnamefont {S.}~\bibnamefont
  {Yamada}}, \bibinfo {author} {\bibfnamefont {K.}~\bibnamefont {Maezawa}},
  \bibinfo {author} {\bibfnamefont {W.~T.}\ \bibnamefont {Yuen}}, \ and\
  \bibinfo {author} {\bibfnamefont {R.~A.}\ \bibnamefont {Stradling}},\
  }\bibfield  {title} {\enquote {\bibinfo {title} {X-conduction-electron
  transport in very thin alas quantum wells},}\ }\href {\doibase
  10.1103/PhysRevB.49.2189} {\bibfield  {journal} {\bibinfo  {journal} {Phys.
  Rev. B}\ }\textbf {\bibinfo {volume} {49}},\ \bibinfo {pages} {2189}
  (\bibinfo {year} {1994})}\BibitemShut {NoStop}%
\bibitem [{\citenamefont {Momose}\ \emph {et~al.}(1999)\citenamefont {Momose},
  \citenamefont {Mori}, \citenamefont {Hamaguchi}, \citenamefont {Ikaida},
  \citenamefont {Arimoto},\ and\ \citenamefont {Miura}}]{Momose1999}%
  \BibitemOpen
  \bibfield  {author} {\bibinfo {author} {\bibfnamefont {H.}~\bibnamefont
  {Momose}}, \bibinfo {author} {\bibfnamefont {N.}~\bibnamefont {Mori}},
  \bibinfo {author} {\bibfnamefont {C.}~\bibnamefont {Hamaguchi}}, \bibinfo
  {author} {\bibfnamefont {T.}~\bibnamefont {Ikaida}}, \bibinfo {author}
  {\bibfnamefont {H.}~\bibnamefont {Arimoto}}, \ and\ \bibinfo {author}
  {\bibfnamefont {N.}~\bibnamefont {Miura}},\ }\bibfield  {title} {\enquote
  {\bibinfo {title} {{Cyclotron resonance in (GaAs)$_{n}$/(AlAs)$_{n}$
  superlattices under ultra-high magnetic fields}},}\ }\href {\doibase
  10.1016/S1386-9477(99)00018-1} {\bibfield  {journal} {\bibinfo  {journal}
  {Physica E}\ }\textbf {\bibinfo {volume} {4}},\ \bibinfo {pages} {286}
  (\bibinfo {year} {1999})}\BibitemShut {NoStop}%
\bibitem [{\citenamefont {Vakili}\ \emph {et~al.}(2004)\citenamefont {Vakili},
  \citenamefont {Shkolnikov}, \citenamefont {Tutuc}, \citenamefont {{De
  Poortere}},\ and\ \citenamefont {Shayegan}}]{Vakili2004}%
  \BibitemOpen
  \bibfield  {author} {\bibinfo {author} {\bibfnamefont {K.}~\bibnamefont
  {Vakili}}, \bibinfo {author} {\bibfnamefont {Y.~P.}\ \bibnamefont
  {Shkolnikov}}, \bibinfo {author} {\bibfnamefont {E.}~\bibnamefont {Tutuc}},
  \bibinfo {author} {\bibfnamefont {E.~P.}\ \bibnamefont {{De Poortere}}}, \
  and\ \bibinfo {author} {\bibfnamefont {M.}~\bibnamefont {Shayegan}},\
  }\bibfield  {title} {\enquote {\bibinfo {title} {{Spin susceptibility of
  two-dimensional electrons in narrow AlAs quantum wells}},}\ }\href {\doibase
  10.1103/PhysRevLett.92.226401} {\bibfield  {journal} {\bibinfo  {journal}
  {Phys. Rev. Lett.}\ }\textbf {\bibinfo {volume} {92}},\ \bibinfo {pages}
  {226401} (\bibinfo {year} {2004})}\BibitemShut {NoStop}%
\bibitem [{\citenamefont {Hirakawa}\ and\ \citenamefont
  {Sakaki}(1986)}]{Hirakawa1986}%
  \BibitemOpen
  \bibfield  {author} {\bibinfo {author} {\bibfnamefont {K.}~\bibnamefont
  {Hirakawa}}\ and\ \bibinfo {author} {\bibfnamefont {H.}~\bibnamefont
  {Sakaki}},\ }\bibfield  {title} {\enquote {\bibinfo {title} {{Mobility of the
  two-dimensional electron gas at selectively doped $n$-type
  Al$_{x}$Ga$_{1-x}$As/GaAs heterojunctions with controlled electron
  concentrations}},}\ }\href {\doibase 10.1103/PhysRevB.33.8291} {\bibfield
  {journal} {\bibinfo  {journal} {Phys. Rev. B}\ }\textbf {\bibinfo {volume}
  {33}},\ \bibinfo {pages} {8291} (\bibinfo {year} {1986})}\BibitemShut
  {NoStop}%
\bibitem [{\citenamefont {Coleridge}(1991)}]{Coleridge1991}%
  \BibitemOpen
  \bibfield  {author} {\bibinfo {author} {\bibfnamefont {P.~T.}\ \bibnamefont
  {Coleridge}},\ }\bibfield  {title} {\enquote {\bibinfo {title} {{Small-angle
  scattering in two-dimensional electron gases}},}\ }\href {\doibase
  10.1103/PhysRevB.44.3793} {\bibfield  {journal} {\bibinfo  {journal} {Phys.
  Rev. B}\ }\textbf {\bibinfo {volume} {44}},\ \bibinfo {pages} {3793}
  (\bibinfo {year} {1991})}\BibitemShut {NoStop}%
\bibitem [{Note1()}]{Note1}%
  \BibitemOpen
  \bibinfo {note} {For the data in the weak screening regime taken with
  $V_\protect \text {FG} \leq -1.7$~V set at room temperature, analysis taking
  into account density inhomogeneity \cite {Qian2017} was necessary to fit the
  Dingle plot with the correct intercept of 4 at $B^{-1}=0$. The density
  inhomogeneity derived from the fit was 0.6 and 1.8\% for $V_\protect \text
  {FG}=-1.70$ and $-1.88$~V, respectively.}\BibitemShut {Stop}%
\bibitem [{\citenamefont {Chen}\ \emph {et~al.}(2012)\citenamefont {Chen},
  \citenamefont {Wang}, \citenamefont {Klochan}, \citenamefont {Micolich},
  \citenamefont {{Das Gupta}}, \citenamefont {Sfigakis}, \citenamefont
  {Ritchie}, \citenamefont {Reuter}, \citenamefont {Wieck},\ and\ \citenamefont
  {Hamilton}}]{Chen2012}%
  \BibitemOpen
  \bibfield  {author} {\bibinfo {author} {\bibfnamefont {J.~C.~H.}\
  \bibnamefont {Chen}}, \bibinfo {author} {\bibfnamefont {D.~Q.}\ \bibnamefont
  {Wang}}, \bibinfo {author} {\bibfnamefont {O.}~\bibnamefont {Klochan}},
  \bibinfo {author} {\bibfnamefont {A.~P.}\ \bibnamefont {Micolich}}, \bibinfo
  {author} {\bibfnamefont {K.}~\bibnamefont {{Das Gupta}}}, \bibinfo {author}
  {\bibfnamefont {F.}~\bibnamefont {Sfigakis}}, \bibinfo {author}
  {\bibfnamefont {D.~A.}\ \bibnamefont {Ritchie}}, \bibinfo {author}
  {\bibfnamefont {D.}~\bibnamefont {Reuter}}, \bibinfo {author} {\bibfnamefont
  {A.~D.}\ \bibnamefont {Wieck}}, \ and\ \bibinfo {author} {\bibfnamefont
  {A.~R.}\ \bibnamefont {Hamilton}},\ }\bibfield  {title} {\enquote {\bibinfo
  {title} {{Fabrication and characterization of ambipolar devices on an undoped
  AlGaAs/GaAs heterostructure}},}\ }\href {\doibase 10.1063/1.3673837}
  {\bibfield  {journal} {\bibinfo  {journal} {Appl. Phys. Lett.}\ }\textbf
  {\bibinfo {volume} {100}},\ \bibinfo {pages} {052101} (\bibinfo {year}
  {2012})}\BibitemShut {NoStop}%
\bibitem [{\citenamefont {Wang}\ \emph {et~al.}(2013)\citenamefont {Wang},
  \citenamefont {Chen}, \citenamefont {Klochan}, \citenamefont {{Das Gupta}},
  \citenamefont {Reuter}, \citenamefont {Wieck}, \citenamefont {Ritchie},\ and\
  \citenamefont {Hamilton}}]{Wang2013}%
  \BibitemOpen
  \bibfield  {author} {\bibinfo {author} {\bibfnamefont {D.~Q.}\ \bibnamefont
  {Wang}}, \bibinfo {author} {\bibfnamefont {J.~C.~H.}\ \bibnamefont {Chen}},
  \bibinfo {author} {\bibfnamefont {O.}~\bibnamefont {Klochan}}, \bibinfo
  {author} {\bibfnamefont {K.}~\bibnamefont {{Das Gupta}}}, \bibinfo {author}
  {\bibfnamefont {D.}~\bibnamefont {Reuter}}, \bibinfo {author} {\bibfnamefont
  {A.~D.}\ \bibnamefont {Wieck}}, \bibinfo {author} {\bibfnamefont {D.~A.}\
  \bibnamefont {Ritchie}}, \ and\ \bibinfo {author} {\bibfnamefont {A.~R.}\
  \bibnamefont {Hamilton}},\ }\bibfield  {title} {\enquote {\bibinfo {title}
  {{Influence of surface states on quantum and transport lifetimes in
  high-quality undoped heterostructures}},}\ }\href {\doibase
  10.1103/PhysRevB.87.195313} {\bibfield  {journal} {\bibinfo  {journal} {Phys.
  Rev. B}\ }\textbf {\bibinfo {volume} {87}},\ \bibinfo {pages} {195313}
  (\bibinfo {year} {2013})}\BibitemShut {NoStop}%
\bibitem [{\citenamefont {Qian}\ \emph {et~al.}(2017)\citenamefont {Qian},
  \citenamefont {Nakamura}, \citenamefont {Fallahi}, \citenamefont {Gardner},
  \citenamefont {Watson}, \citenamefont {L{\"{u}}scher}, \citenamefont {Folk},
  \citenamefont {Cs{\'{a}}thy},\ and\ \citenamefont {Manfra}}]{Qian2017}%
  \BibitemOpen
  \bibfield  {author} {\bibinfo {author} {\bibfnamefont {Q.}~\bibnamefont
  {Qian}}, \bibinfo {author} {\bibfnamefont {J.}~\bibnamefont {Nakamura}},
  \bibinfo {author} {\bibfnamefont {S.}~\bibnamefont {Fallahi}}, \bibinfo
  {author} {\bibfnamefont {G.~C.}\ \bibnamefont {Gardner}}, \bibinfo {author}
  {\bibfnamefont {J.~D.}\ \bibnamefont {Watson}}, \bibinfo {author}
  {\bibfnamefont {S.}~\bibnamefont {L{\"{u}}scher}}, \bibinfo {author}
  {\bibfnamefont {J.~A.}\ \bibnamefont {Folk}}, \bibinfo {author}
  {\bibfnamefont {G.~A.}\ \bibnamefont {Cs{\'{a}}thy}}, \ and\ \bibinfo
  {author} {\bibfnamefont {M.~J.}\ \bibnamefont {Manfra}},\ }\bibfield  {title}
  {\enquote {\bibinfo {title} {{Quantum lifetime in ultrahigh quality GaAs
  quantum wells: Relationship to $\Delta_{5/2}$ and impact of density
  fluctuations}},}\ }\href {\doibase 10.1103/PhysRevB.96.035309} {\bibfield
  {journal} {\bibinfo  {journal} {Phys. Rev. B}\ }\textbf {\bibinfo {volume}
  {96}},\ \bibinfo {pages} {035309} (\bibinfo {year} {2017})}\BibitemShut
  {NoStop}%
\bibitem [{Note2()}]{Note2}%
  \BibitemOpen
  \bibinfo {note} {An attempt to fit the Dingle-plot data in Fig.~\ref
  {Fig5}(a) using the density-gradient model in Ref. \cite {Qian2017} together
  with the calculated quantum lifetime resulted in a strongly nonlinear curve,
  which did not fit the experimental results.}\BibitemShut {Stop}%
\bibitem [{Note3()}]{Note3}%
  \BibitemOpen
  \bibinfo {note} {Reducing the field sweep rate from 10 mT/sec (which we
  normally use) to 10 $\mu $T/sec did not affect the measured value of $\mu
  _\protect \text {q}$.}\BibitemShut {Stop}%
\bibitem [{\citenamefont {Fu}\ \emph {et~al.}(2018)\citenamefont {Fu},
  \citenamefont {Riedl}, \citenamefont {Borisov}, \citenamefont {Zudov},
  \citenamefont {Watson}, \citenamefont {Gardner}, \citenamefont {Manfra},
  \citenamefont {Baldwin}, \citenamefont {Pfeiffer},\ and\ \citenamefont
  {West}}]{Fu2018}%
  \BibitemOpen
  \bibfield  {author} {\bibinfo {author} {\bibfnamefont {X.}~\bibnamefont
  {Fu}}, \bibinfo {author} {\bibfnamefont {A.}~\bibnamefont {Riedl}}, \bibinfo
  {author} {\bibfnamefont {M.}~\bibnamefont {Borisov}}, \bibinfo {author}
  {\bibfnamefont {M.~A.}\ \bibnamefont {Zudov}}, \bibinfo {author}
  {\bibfnamefont {J.~D.}\ \bibnamefont {Watson}}, \bibinfo {author}
  {\bibfnamefont {G.}~\bibnamefont {Gardner}}, \bibinfo {author} {\bibfnamefont
  {M.~J.}\ \bibnamefont {Manfra}}, \bibinfo {author} {\bibfnamefont {K.~W.}\
  \bibnamefont {Baldwin}}, \bibinfo {author} {\bibfnamefont {L.~N.}\
  \bibnamefont {Pfeiffer}}, \ and\ \bibinfo {author} {\bibfnamefont {K.~W.}\
  \bibnamefont {West}},\ }\bibfield  {title} {\enquote {\bibinfo {title}
  {{Effect of illumination on quantum lifetime in GaAs quantum wells}},}\
  }\href {\doibase 10.1103/PhysRevB.98.195403} {\bibfield  {journal} {\bibinfo
  {journal} {Phys. Rev. B}\ }\textbf {\bibinfo {volume} {98}},\ \bibinfo
  {pages} {195403} (\bibinfo {year} {2018})}\BibitemShut {NoStop}%
\bibitem [{\citenamefont {Kang}\ \emph {et~al.}(1995)\citenamefont {Kang},
  \citenamefont {He}, \citenamefont {Stormer}, \citenamefont {Pfeiffer},
  \citenamefont {Baldwin},\ and\ \citenamefont {West}}]{Kang1995}%
  \BibitemOpen
  \bibfield  {author} {\bibinfo {author} {\bibfnamefont {W.}~\bibnamefont
  {Kang}}, \bibinfo {author} {\bibfnamefont {Song}\ \bibnamefont {He}},
  \bibinfo {author} {\bibfnamefont {H.~L.}\ \bibnamefont {Stormer}}, \bibinfo
  {author} {\bibfnamefont {L.~N.}\ \bibnamefont {Pfeiffer}}, \bibinfo {author}
  {\bibfnamefont {K.~W.}\ \bibnamefont {Baldwin}}, \ and\ \bibinfo {author}
  {\bibfnamefont {K.~W.}\ \bibnamefont {West}},\ }\bibfield  {title} {\enquote
  {\bibinfo {title} {{Temperature dependent scattering of composite
  fermions}},}\ }\href {\doibase 10.1103/PhysRevLett.75.4106} {\bibfield
  {journal} {\bibinfo  {journal} {Phys. Rev. Lett.}\ }\textbf {\bibinfo
  {volume} {75}},\ \bibinfo {pages} {4106} (\bibinfo {year}
  {1995})}\BibitemShut {NoStop}%
\end{thebibliography}%
%

\end{document}